\documentclass[aps,showpacs,twocolumn,superscriptaddress,nofootinbib,floatfix,section 10pt]{revtex4-2}
\usepackage{amsfonts,amssymb,stmaryrd,latexsym,amsmath,braket}
\usepackage{graphicx,subfigure}
\usepackage{comment}
\usepackage{times}
\usepackage{slashed}
\usepackage{bm}
\usepackage{braket}
\usepackage{appendix}
\usepackage{enumitem}
\usepackage{multirow}
\usepackage{array}

\usepackage[colorlinks=true,backref=true,linktocpage=true,citecolor=blue,urlcolor=blue,linkcolor=blue,pdfpagemode=UseOutlines]{hyperref}
\usepackage[dvipsnames]{xcolor}

\begin{document}
\title{Numerical study of computational cost of maintaining adiabaticity for long paths}
\author{Thomas D. Cohen}
\email{cohen@umd.edu}
\affiliation{Department of Physics and Maryland Center for Fundamental Physics, University of Maryland, College Park, MD 20742 USA}

\author{Hyunwoo Oh}
\email{hyunwooh@umd.edu}
\affiliation{Department of Physics and Maryland Center for Fundamental Physics, University of Maryland, College Park, MD 20742 USA}

\author{Veronica Wang}
\email{veronicawang2007@gmail.com}
\affiliation{Poolesville High School, Poolesville, MD 20837 USA}

\begin{abstract}

Recent work~\cite{Cohen:2024nbk} argued that the scaling of a dimensionless quantity $Q_D$ with path length is a better proxy for quantifying the scaling of the computational cost of maintaining adiabaticity than the timescale. It also conjectured that generically the scaling will be superlinear (although special cases exist in which it is linear). The quantity $Q_D$ depends only on the properties of ground states along the Hamiltonian path and the rate at which the path is followed.  In this paper, we demonstrate that this conjecture holds for simple Hamiltonian systems that can be studied numerically. In particular, the systems studied exhibit the behavior that $Q_D$ grows approximately as $L \log L$ where $L$ is the path length when the threshold error is fixed.
\end{abstract}

\date{\today}
\maketitle

\section{Introduction} \label{Sec:Introduction}

Simulating quantum physics on classical computers has been enormously successful in many areas of physics but it is restricted by the notorious sign problem. Quantum computation is a candidate to avoid the sign problem entirely by using Hamiltonian mechanics. The basic process of simulations for physics on quantum processors involves three steps: preparing an initial state, evolving the state using a given Hamiltonian, and measuring the final state. 

In general, it is difficult to prepare ground states of arbitrary Hamiltonians on quantum computers. The adiabatic theorem is one possible way to prepare the initial state by exploiting the fact that if the system changes infinitely slowly, then the eigenstates of the initial Hamiltonian flow to the corresponding eigenstates of the final Hamiltonian. Therefore, if one can prepare the ground state of an initial Hamiltonian with the knowledge of that system, the desired ground state can be prepared through the adiabatic evolution.

In practice, the system cannot evolve infinitely slowly, so errors cannot be avoided from these diabatic evolutions. Therefore, methods for estimating errors are important and have been extensively investigated. In general, there are two directions to estimate errors, using the ``switching theorem''~\cite{LENARD1959261, GARRIDO1962553,10.1063/1.1704127, Nenciu1981, Nenciu1991, Nenciu1993, HAGEDORN2002235, 10.1063/1.4748968} and finding sufficient conditions for having errors smaller than some threshold error using the norm of derivatives of a given Hamiltonian and its gap~\cite{doi:10.1143/JPSJ.5.435, 10.1063/1.2798382, messiah61, PhysRevA.80.012106, Cheung_2011, Mozgunov:2020cof, Burgarth2022oneboundtorulethem}. 

First, the switching theorem is a statement that the errors follow the asymptotic series in the timescale and each term in the series only depends on the initial and the final Hamiltonians and how the system leaves the initial Hamiltonian and arrives at the final Hamiltonian. This theorem estimates errors, as opposed to mere bounds for errors, via the asymptotic series and becomes accurate in the limit of asymptotically slow evolution. However, it is often impractical to estimate errors this way as the asymptotic regime in which the series holds can require extremely large timescales~\cite{Cohen:2025lij}.

The other direction is to find sufficient conditions using the norm of derivatives of the Hamiltonian and its gap, which is usually expressed as an inequality for the timescale and such rigorous bounds depend on the value of the maximum eigenvalue. However, these bounds are typically of little, if any, utility for many-body theory, especially for quantum field theories. This is because in order for the theory to be simulated on finite-qubit quantum simulators, it must be regulated to have a finite Hilbert space---ideally in a way that does not significantly affect the physical results. The regulation necessarily affects the highest eigenvalue of the Hamiltonian and its derivatives, which in turn also affects the bounds on the errors. However, if one prepares the ground state, the errors are mainly dependent on the properties of the ground state and the low-lying states. Therefore, as the regulator gets large, pushing the system closer to the physical system of interest, these bounds become weaker and give an extreme overestimate for the timescales for which the errors are within some threshold.

Additionally, the timescale for quantifying the cost of maintaining adiabaticity itself is problematic: from the uncertainty principle, an increase in the energy scale results in a corresponding decrease in the timescale. The scaling of computational cost of maintaining a fixed error for systems in which the typical energy scale varies along is not well reflected by using the total time as a reflection of the cost.

Recently, Ref.~\cite{Cohen:2024nbk} suggested that there is a class of dimensionless quantity, $Q_D$, that reflects the cost of maintaining adiabaticity better than the timescale. That work conjectured that such a quantity will generally grow superlinearly (faster than linearly) with respect to the path length in the large length limit, although special cases exist where the scaling is linear. 

The long length limit is important because for simulating large systems, such as quantum field theories~\cite{Jordan:2012xnu, Jordan:2011ci, Jordan:2014tma, Liu:2020eoa, Chakraborty:2020uhf, Buser:2020uzs, Gharibyan:2020bab, Kreshchuk:2020dla, Honda:2021aum, Ciavarella:2022qdx, Li:2022ped, Turco:2023rmx, Farrell:2024fit, Kaikov:2024acw, Li:2024lrl, Lee:2024jnt, DAnna:2024mmz}, the path length from the initial Hamiltonian to the final Hamiltonian tends to be large. For example, Ref.~\cite{Cohen:2023dll} shows that for local field theories, the path length is proportional to the square-root of the volume of the system. Moreover, Ref.~\cite{Cohen:2023dll} shows that methods of state preparation that are known to scale linearly with the path length exist; thus, to determine whether adiabatic state preparation is competitive for long path lengths, it is important to know how the cost of adiabatic state preparation scales.

The goal of this paper is to provide numerical evidence in support of the conjecture in~\cite{Cohen:2024nbk} by explicitly calculating the scaling for relatively simple Hamiltonians that can be simulated accurately on classical computers. 

The computational cost---estimated by whatever proxy is used---will depend on how the path through Hamiltonian space is traversed and the information about what is happening during the traversal. In certain settings, finding the state of interest is equivalent to solving the problem at hand. For such problems, one only needs to traverse the path successfully only once and hence one cannot get significant information about what occurs during successful traversals until after it is too late to be useful. In contrast, in many physics applications one needs to produce the same state repeatedly in order to study its properties. This paper focuses on the latter situation, where significant information about the traversal has been obtained via previous successful traversals: the proxies we use for computational cost depend on this information. Moreover, it is assumed that the actual cost can closely follow the proxies once information about the proxies is extracted.

This paper is structured as follows: In Section~\ref{Sec:Background}, the basic concepts behind the conjecture in~\cite{Cohen:2024nbk} are introduced. The numerical methods to test the conjecture are discussed in Section~\ref{Sec:Methods}. The results are presented in Section~\ref{Sec:Results}. The implications are discussed in Section~\ref{Sec:Discussion}.

\section{Background} \label{Sec:Background}

This section introduces some basic quantities including those that serve as proxies for the cost of maintaining adiabaticity of an evolving system. These are needed in order to formulate the conjecture in~\cite{Cohen:2024nbk}.

\subsection{Errors}

Since one cannot evolve quantum systems infinitely slowly, diabatic errors in adiabatic state preparation are unavoidable. If one define $U(t_f, t_i)$ as a time evolution operator from the initial Hamiltonian to the final Hamiltonian, the error, $\epsilon$, is defined as
\begin{equation}
\epsilon \equiv \left \lVert \left (1-|g_f\rangle \langle g_f |  \right) U(t_f,t_i)  |g_i \rangle \right \rVert .
   \label{Eq:errordef}
\end{equation}
Note that the error is defined as a magnitude and is therefore always positive. While this paper focuses on the preparation of ground states, the definition of error can be generalized to any state in a straightforward manner without loss of generality.

\subsection{Path length}

Our goal is to describe the scaling behavior with respect to path length, making it essential to first define what we mean by path length. One can define a dimensionless path length for each trajectory~\cite{10.5555/2011804.2011811, boixo2010fast, PhysRevA.89.012314, Cohen:2023dll}. If one consider a Hamiltonian path parametrized by an arbitrary variable, $s$, the path length is defined as
\begin{align}
L_{a,b} & \equiv \int_{s_a}^{s_b}  d s \, \lVert   | g'(s) \rangle -| g(s) \rangle \langle g(s)| g'(s) \rangle  \rVert,  \label{Eq:L0} \\
{\rm with} \; & \; | g'(s) \rangle  \equiv \frac{d}{d s} |g(s)\rangle. \label{Eq:g'(s)}
\end{align} 
Note that the definition of path length is independent of the parametrization of the Hamiltonian. One can simplify the path length formula by exploiting the fact that the phase of the ground state at any point along the path is arbitrary. If one chooses a phase such that the integrand of the Berry phase~\cite{Berry:1984jv}, $\langle g(s) | g'(s) \rangle$, is zero along the path, then the path length becomes
\begin{equation}
L_{a,b}  = \int_{s_a}^{s_b}  d s \, \lVert  | g'(s) \rangle  \rVert.  \label{Eq:L} 
\end{equation}

It is often useful in formal analysis to parametrize a path in terms of the path length covered starting at the initial point:
\begin{equation}
    \lambda(s) \equiv \lambda_i + \int_{s_i}^s d \tilde{s} \; \lVert  |g'(\tilde{s})\rangle   \rVert .
\label{Eq:lambda}
\end{equation}
With this parametrization, $L_{a,b}  = \lambda_b-\lambda_a$ and the Schr\"odinger equation becomes
\begin{equation}
    i \frac{d}{d\lambda} | \psi \rangle  = \frac{H(\lambda)}{v(\lambda)} | \psi \rangle, \label{Eq:SElambda}
\end{equation}
where $v(\lambda) = \frac{d \lambda}{dt}$ is the velocity along the path.

Reference~\cite{Cohen:2024nbk} showed that time is not, in general, a good proxy for quantifying the scaling of adiabaticity with path length. It proved a no-go theorem and proposed alternative quantities that better represent computational cost to achieve a fixed small error. In the following subsections, we will review this claim. For a detailed explanation, readers may refer to the appendix of the original work~\cite{Cohen:2024nbk}.

\subsection{A no-go theorem}

A no-go theorem in~\cite{Cohen:2024nbk} implies that two Hamiltonians $H_A(t)$ and $H_B(t)$ can have the same errors with respect to the path length if they are related by
\begin{equation}
    \frac{H_A(\lambda)}{v_A(\lambda)} = \frac{H_B(\lambda)}{v_B(\lambda)}. 
    \label{Eq:no-go}
\end{equation}

It is trivial from the observation of Eq.~(\ref{Eq:SElambda}) that the two Hamiltonians actually describe the same physics. It is from the observation that if one rescales the Hamiltonian along the path, then one can rescale its velocity by the same factor so that the errors stay the same. This demonstrates why time is not a useful measure for quantifying the scaling of the cost to maintain adiabaticity with path length: time changes under rescaling of the Hamiltonian along the path, while the errors remain the same. Therefore, another quantity should be used to quantify the cost of maintaining adiabaticity in this context.


\subsection{Proxies for the computational cost}


Given the no-go theorem it is useful to construct proxies for the cost to maintain adiabaticity that are better suited to the problem. Such quantities will be denoted by $Q_D$. The issue under consideration is the scaling of the quantity with path length in the regime where errors are 
small. Once a useful proxy is identified, the question of interest becomes how does the quantity scale with path length, subject to the constraint that the error is held fixed.

In addition to being connected with the computational resources needed for a calculation,  reference~\cite{Cohen:2024nbk} identified some reasonable conditions that a useful proxy should satisfy, namely:

\begin{enumerate}
    \item $Q_D$  should be dimensionless. \label{cond1}
    
    \item  $Q_D$  should be the same for two systems related by Eq.~(\ref{Eq:no-go}).  \label{cond2}
    
    \item $Q_D$  should reflect how close the system is to the adiabatic limit in the sense that for a path of fixed length and a fixed relative velocity along the path, $Q_D$ monotonically increases with increasing time. \label{cond3}
    
    \item $Q_D$ should scale superlinearly in $L$ a given fixed error for time-periodic Hamiltonians.   \label{cond4}
\end{enumerate}


The first condition is imposed because meaningful physical results should be expressed in terms of dimensionless quantities.
The second condition is necessary to ensure that the argument leading to the no-go theorem does not apply to $Q_D$ in the same way that it applies to time. 
The third condition is essential for any sensible proxy for the cost of maintaining adiabaticity; this is required since the analysis of cost is in the context of adiabatic quantum computing. 
The fourth condition was observed to hold when using time as a measure of cost in~\cite{Cohen:2024nbk}.\footnote{Note that for time-periodic Hamiltonians, the concerns raised by the no-go theorem do not apply, as there is no secular growth in the scale of the energy.} The result is intuitive in the sense that errors accumulate as the system travels many cycles, therefore in order to have a fixed error while the path length increases, $Q_D$ increases superlinearly in $L$. Any reasonable proxy for the cost of maintaining adiabaticity should behave in the same way. 
Note that the order of the limits for a threshold error and path length is important: the situation considered in this paper is to have a small threshold error to stay in the adiabatic limit while increasing the path length.

\subsection{Quantities reflecting computational costs}

Reference~\cite{Cohen:2024nbk} suggested that the quantity
\begin{align}
    Q_D  & \equiv \int_{\lambda_i}^{\lambda_f} d\lambda \sum_{k\neq g} \left( \frac{E_k(\lambda) - E_g(\lambda)}{v(\lambda)}  \left| \langle k(\lambda)  | g'(\lambda) \rangle \right|^2 \right) \nonumber \\
    & = \int_{\lambda_i}^{\lambda_f} d \lambda  \, \langle g'| \frac{H(\lambda) -E_g(\lambda)}{v(\lambda)} |g' \rangle  \label{Eq:QD1}
\end{align}
is a natural choice for a proxy for computational cost. Note that $Q_D$ varies inversely with velocity; that is, faster traversal results in a lower cost, as expected in any reasonable estimate of computational cost. Additionally, $Q_D$ depends on the energy scale through the energy differences between the ground state and excited states, weighted by the probability that the changing ground state overlaps with those states.  

The reasoning behind this is straightforward: to construct a meaningful dimensionless cost measure, a quantity with the dimension of energy is required to absorb the inverse energy scale associated with $1/v(\lambda)$. The weighted average of the energy differences in the transitions provides a natural energy scale. Furthermore, weighting by the probability of overlap with the changing ground state ensures that the energy scale used is the one most relevant for producing errors.

Thus, any reasonable optimization of computational resources---while maintaining a fixed error threshold---should focus on states within this energy range. Here, computational resources refer to those necessary for performing the calculation, such as the number of gate operations in the context of digital quantum computing, rather than the total number of qubits required. Accordingly, $Q_D$ serves as a dimensionless, albeit somewhat crude, measure of the optimized computational cost.

It is easy to check that $Q_D$ satisfies the conditions enumerated above. 

Note that this quantity is not unique; it is straightforward to construct variations of Eq.~(\ref{Eq:QD1}) that also satisfy the conditions. For example, 
\begin{subequations}
\begin{align}
    Q_{D_2} &\equiv \int_{\lambda_i}^{\lambda_f} d \lambda  \, \frac{\sqrt{\langle g'| \left(H(\lambda) -E_g(\lambda)\right)^2 |g' \rangle }} {v(\lambda)}  \label{Eq:QD2}, \   {\rm and} \\
    Q_{D_{1/2}} &\equiv \int_{\lambda_i}^{\lambda_f} d \lambda  \, \frac{\left ( \langle g'| \sqrt {H(\lambda) -E_g(\lambda) } |g' \rangle \right)^2} {v(\lambda)}  \label{Eq:QD3}
\end{align}
both satisfy the conditions.

$Q_{D_2}$ prioritizes contributions from high-energy states while $Q_{D_{1/2}}$ emphasizes contributions from the low-lying states. In general, anything of the form 
\begin{equation}
Q_{D_f} \equiv \int_{\lambda_i}^{\lambda_f} d \lambda  \, \frac{f^{-1} \left(\langle g'| f\left(H(\lambda) -E_g(\lambda)\right) |g' \rangle \right)} {v(\lambda)}  \label{Eq:QD4}
\end{equation}
for any monotonically increasing function $f$ will satisfy the conditions.  

Any of these provides a reasonable, albeit somewhat qualitative, measure of the computational resources required to maintain adiabaticity.

\end{subequations}

\subsection{Conjecture} 

The conjecture of Ref.~\cite{Cohen:2024nbk} is that the computational cost as measured by any of the proxies listed above will generically scale superlinearly with the path length (in the asymptotic limit) for trajectories in which the total error is held to be small and fixed, although special cases exist where the scaling is linear.

The motivation for this conjecture is simple and related to the condition~\ref{cond4} above. That condition was shown to hold for time-periodic Hamiltonians, using time as a measure of the cost to maintain adiabaticity~\cite{Cohen:2024nbk}. In those cases there is no secular growth in the scale of the Hamiltonians so the concerns associated with the no-go theorem are not germane. This result is intuitive since errors accumulate as the system undergoes multiple cycles. Therefore, to maintain a fixed error while increasing the path length, the system must slow down, leading to a superlinear increase in the total time with respect to $L$. Any sensible proxy for measuring the cost to maintain adiabaticity should behave in the same way. It is plausible that general Hamiltonians that are not periodic in time should behave similarly due to an analogous argument: when the timescale is fixed, errors accumulate as the system travels longer, so in order to have a fixed error, the rate of the change of the system should be slowed down. Therefore one can conjecture that just as for periodic Hamiltonians the useful proxies for cost should scale superlinearly in $L$.

The goal of this paper is to test this conjecture via the study of simple systems that can be simulated accurately for long paths with present computers.

\section{Approach for numerical tests} \label{Sec:Methods}

This section describes the approach to set up numerical tests of the conjecture 
for various time-dependent Hamiltonians. Note that the methods described here are for the purpose of testing the scaling of the proxies for the computational cost of maintaining adiabaticity with path length; they do not describe how one would implement adiabatic calculations in realistic settings.

The conjecture has been stated in terms of $\lambda$ as the parametrization for path length; this is the most transparent parametrization theoretically. However, it is technically complicated to evolve a system numerically directly in terms of this parametrization. Therefore, it is useful to do the numerical evolution of the state using time to specify where the system is along the path.

In doing this it is important to recall that to keep errors below a fixed threshold error, the velocity $v(\lambda)$, which characterizes the traversal of the path, needs to decrease as the path length increases. In this paper, we consider a family of velocities with the same relative velocities but different absolute velocities. The velocity can be parametrized with a scaling factor $s_c$:
\begin{equation}
    v(\lambda) = \frac{v_{\rm ref}(\lambda)}{s_c},
    \label{Eq:scale}
\end{equation}
where $v_{\rm ref}$ is a reference velocity. $s_c$ is dimensionless but characterizes the overall timescale of the traversal of the path at fixed path length. Of course, $s_c$ can be chosen as any value and any sensible accounting for computational costs will decrease with increasing $s_c$. However, one cannot choose $s_c$ to be arbitrarily large while keeping the error bounded by the fixed threshold error. The goal is to minimize computational costs while meeting the threshold error condition. Thus it makes sense to fix the value of $s_c$ to be the value that yields an error matching the threshold\footnote{There is a subtlety in this: as $s_c$ is decreased from some initial value, the error need not increase monotonically. It can oscillate around an increasing function, yielding multiple values of $s_c$ for which the error matches the threshold. The issue then becomes which of these values is the most relevant. If the goal is to find the absolute minimum cost, then clearly the smallest value of $s_c$ satisfying the condition is the correct choice. However, such a choice is problematic if the goal is to reliably ensure that the error is bounded: in realistic scenarios in which $s_c$ cannot be arbitrarily fine-tuned, small deviations from the optimal choice can lead to much larger error. Thus, this paper adopts the more conservative strategy of picking the largest value that yields the threshold error. \label{footnote:sc}}. Thus, $s_c$ chosen will depend on the path and the threshold error, $\epsilon_{\rm th}$: $s_c=s_c(\epsilon_{\rm th}, L; H) $.  

With $s_c$ fixed, we can re-express our proxies in terms of time so that the conjectured proxies for quantifying adiabaticity. For example, using Eq.~(\ref{Eq:lambda}) and the chain rule, Eq.~(\ref{Eq:QD1}) becomes
\begin{equation}
\begin{aligned}
    Q_D &= \int_{\lambda_i}^{\lambda_f} d\lambda \, \langle \frac{dg}{d\lambda} |\frac{  H(\lambda)-E_g(\lambda) }{v(\lambda)} | \frac{dg}{d\lambda} \rangle \\
    & = \int_{0}^T dt \, \frac{d\lambda}{dt}  \frac{dt}{d\lambda} \langle \frac{dg}{dt}| \frac{H(t) - E_g(t)}{d\lambda/dt} | \frac{dg}{dt} \rangle \frac{dt}{d\lambda} \\
    & = \int_0^T dt \, \langle \dot{g} | \frac{H_(t)-E_g(t)}{(d\lambda/dt)^2} | \dot{g} \rangle \\
    & = \int_0^T dt \, \langle \dot{g} | \frac{H_(t)-E_g(t)}{\langle \dot{g}| \dot{g} \rangle} | \dot{g} \rangle.
\end{aligned}
\end{equation}
Therefore, Eqs.~(\ref{Eq:QD1}),~(\ref{Eq:QD2}), and~(\ref{Eq:QD3}), become
\begin{widetext}
\begin{subequations}
\begin{align}
    Q_{D_1}(\epsilon_{\rm th}, T_{\rm end}) & =  s_c(\epsilon_{\rm th}, T_{\rm end}) \int_0^{T_{\rm end}} dt \; \frac{\langle \dot{g} | \left( H(t) - E_g(t) \right) | \dot{g} \rangle}{\langle \dot{g} | \dot{g} \rangle}  \\    
    Q_{D_2}(\epsilon_{\rm th}, T_{\rm end}) & = s_c(\epsilon_{\rm th}, T_{\rm end}) \int_0^{T_{\rm end}} dt \; \frac{ \sqrt{\langle \dot{g} | \left( H(t) - E_g(t) \right)^2 | \dot{g} \rangle}}{ \sqrt{\langle \dot{g} | \dot{g} \rangle}}  \\
    Q_{D_{1/2}}(\epsilon_{\rm th}, T_{\rm end}) & = s_c (\epsilon_{\rm th}, T_{\rm end}) \int_0^{T_{\rm end}} dt \; \frac{ \left(\langle \dot{g} | \sqrt{ H(t) - E_g(t) } | \dot{g} \rangle \right)^2}{\left|\langle \dot{g} | \dot{g} \rangle \right|^2}  
\end{align} \label{Eq:Qds}
\end{subequations}
\end{widetext}
Since we parameterize with time, $T_{\rm end}$, the time at the end of the path parametrizes how far along the path is probed. 

The basic strategy to estimate $Q_D$ for different path lengths is has several steps:
\begin{enumerate}
    \item Find a scale factor $s_c$ for a given Hamiltonian with $T_{\rm end}$ that corresponds to the path length of interest $L$. This is done by starting from a scale factor $s'$ known to be too large to have very small errors and decreasing it until the error by solving the Schr\"odinger equation reaches the threshold error $\epsilon_{\rm th}$.  This will yield the appropriate scale factor as discussed in footnote~\ref{footnote:sc}.  One can calculate $Q_D$ for that end point
    \item Increase the path length to $L+dL$ by multiplying a constant factor k to $T_{\rm end}$: $T'_{\rm end} = k T_{\rm end}$, where $k>1$. Choose $s'$ as $\kappa s_c$ where $\kappa$ is another constant factor larger than 1\footnote{Note that $k$ and $\kappa$ are tuning parameters for algorithmic performance rather than physical variables}.
    \item Go back to the step 1.
\end{enumerate}
One can then study how $Q_D$ scales with $L$ along the chosen path with fixed relative velocities.

One might worry that this approach uses the dimensionful parameter of time, while the analysis in the previous section is based on the path length and quantities related to it. However, all relevant quantities can be calculated this way faithfully reproduce those calculated in terms of the path length since this is just a reparametrization. This reparametrization is chosen for reasons of computational simplicity in this context.


\section{Results} \label{Sec:Results}

This section describes the calculations done to verify the conjectures. To this there is a need to study systems that were simple enough to numerically calculate accurately for long paths. For this we choose low-dimensional Hamiltonians, restricting our attention 4-dimensional Hilbert space so that the Hamiltonians are represented by $4 \times 4$ dimensional matrices. 
Four-dimensional Hilbert spaces are more general than two-dimensional ones, as they allow access to multiple excited states in transitions from the ground state. Moreover, since errors and other relevant quantities, such as path length or $Q_D$, primarily depend on the ground state and low-lying excitations, it is unlikely that the generic behavior is strongly dependent on the Hilbert space dimension. Thus, small-dimensional spaces should provide accurate insights for such studies. 

There are two ways to achieve long path lengths: evolving a small system for an extended duration or working with a larger system. The computational complexity for solving coupled differential equations is $O(N^2)$ for explicit solvers and higher for implicit solvers, where $N$ is the dimension of the Hilbert space. While the path length tends to increase linearly with respect to the end time $T_{\rm end}$ for a given Hamiltonian, the cost of solving differential equations increases $N^2$ with respect to the dimension of the Hamiltonian. Therefore, the optimal approach is to choose a low-dimensional model that is general enough to capture relevant behavior while avoiding overly specific cases and evolve it for a longer time.

\begin{figure}[t]
    \includegraphics[width=0.49\textwidth]{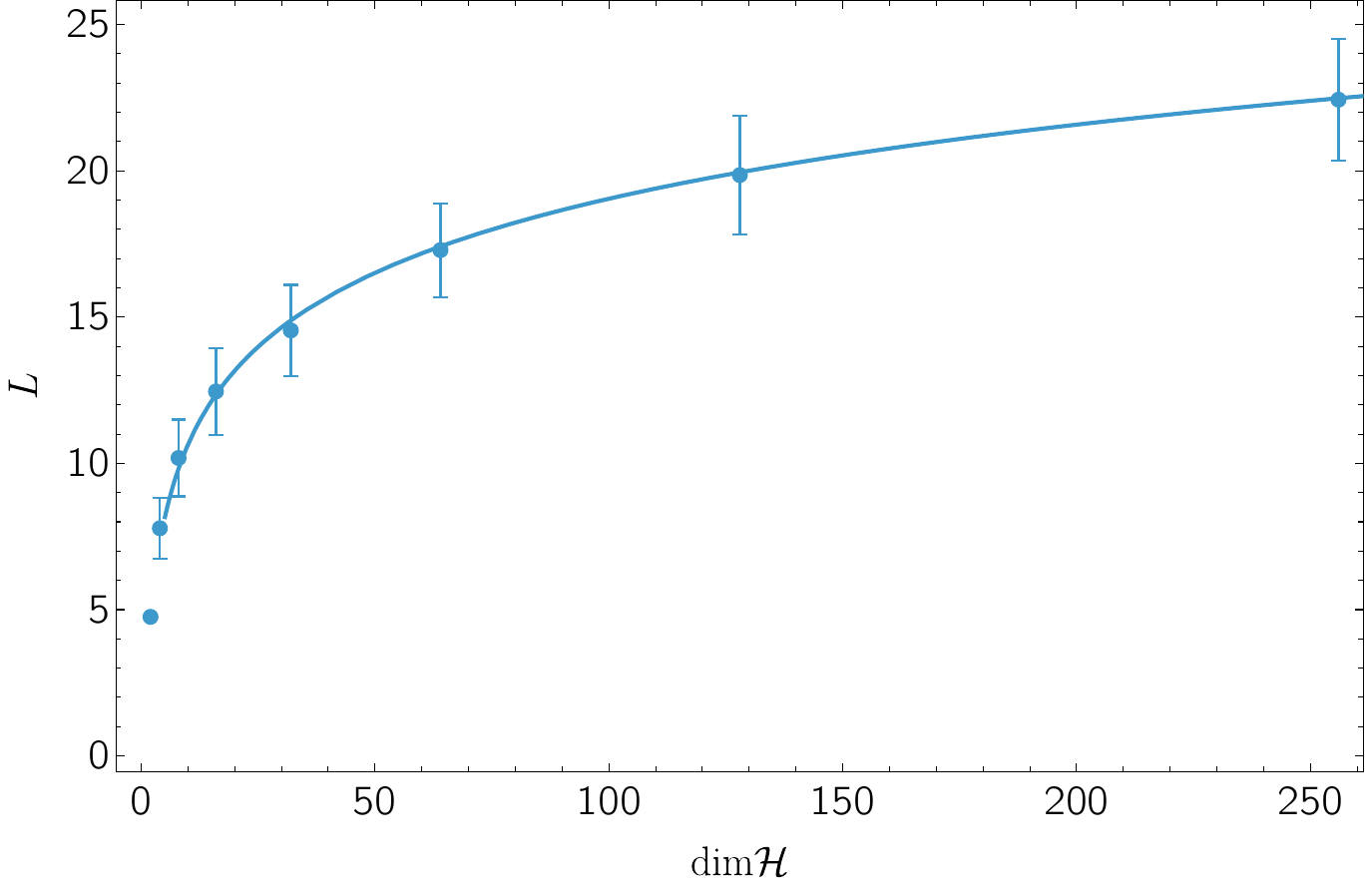}
    
    \caption{Comparison of the path length with different dimensions of Hamiltonians at fixed end time $T_{\rm end}=10$. 100 Hamiltonians are randomly chosen for each dimension in order to estimate means and standard deviations. The form of Hamiltonians are given by Eq.~(\ref{Eq:RandomH}). 
    }

    \label{fig:RH}
\centering 
\end{figure}

Fig.~\ref{fig:RH} shows how the path length depends on the Hilbert space dimension, along with a logarithmic fit: $a \log L+b$. The figure illustrates that the path length increases more slowly as the dimension increases. If we assume this corresponds to a spin-$\frac 1 2$ system, the path length scales proportionally to the system volume, which grows faster than in local field theories~\cite{Cohen:2023dll}. Consequently, an efficient strategy for studying long path length behavior of diabatic errors, especially on classical computers, is to use models with a small Hilbert space dimension.

Note that while numerical differences between the true ground state for the final Hamiltonian and the states produced by slowly evolving the ground state of the initial Hamiltonian can develop numerical errors due to the lack of numerical accuracy in the evolution, for the systems studied here, the dominant error is due to diabatic corrections to the adiabatic theorem, even for very long simulations.

\subsection{Random Hamiltonians}

These studies focus on understanding generic behavior. Therefore, we will study Hamiltonians that are largely chosen randomly within a constrained form. The virtue of using random Hamiltonians is that they are likely to reflect generic features in typical systems without being influenced by the specific properties of particular systems, which may have unusual features that are not usually present and introduce biases in the study of generic behavior. 

\begin{figure}[t]
    \includegraphics[width=0.49\textwidth]{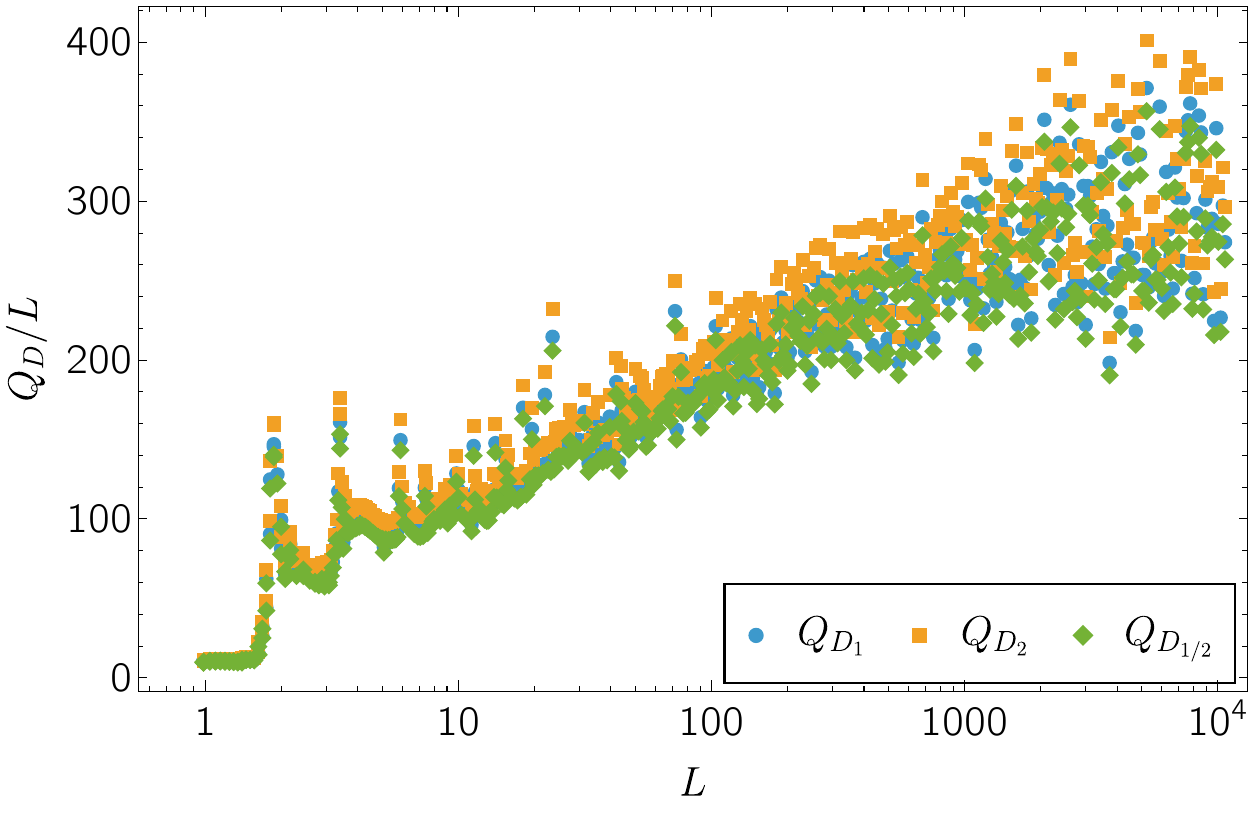}
    
    \caption{ Comparison of different $Q_D$'s suggested in Eq.~(\ref{Eq:Qds}) for a Hamiltonian with Eq.~(\ref{Eq:RandomH}).
    }

    \label{fig:Qs}
\centering 
\end{figure}

The form was chosen to satisfy a few basic conditions:
\begin{enumerate}
    \item The time dependence of the Hamiltonian was guaranteed not to be periodic.  \label{c1}
    \item The gaps in the spectrum change over time, but the various gaps between the ground state and the three excited states in the system are all of the same overall scale.\label{c2}
    \item The Hamiltonian is real and symmetric corresponding to a time-reversal symmetric system. \label{c4}
\end{enumerate}
Condition~\ref{c1} is imposed because the periodic behavior is special and may not reflect generic behavior. In any case, it was already shown in Ref.~\cite{Cohen:2024nbk} that the scaling of the error in time-periodic Hamiltonians with fixed error requires times that grow superlinearly with path length, which will be reflected in the proxies under consideration here. Condition~\ref{c2} is designed to ensure that the system is generic and is not tuned, for example, to have an excited state that is always anomalously low and dominates the error. 
Condition~\ref{c4} reflects our ultimate interest in time-reversal systems; studies similar to the ones done here can be done with general Hermitian Hamiltonians, but then one has to decide separately on the scale for the real-symmetric and imaginary-antisymmetric pieces.

The form of random Hamiltonians that was chosen for these studies was:
\begin{equation}
\begin{aligned}
    H(t) & = \left(2.1\sin(t)  + \sin(\sqrt{2}t)\right) \times 
    H_1 \\
    & + \left(2.7\cos(t) + \cos( \sqrt{2} t) \right) \times H_2,
\end{aligned} \label{Eq:RandomH}
\end{equation}
where $H_1$ and $H_2$ $4 \times 4$ real symmetric matrices whose elements were chose randomly via a normal distribution centered on zero with a standard deviation of unity.  These forms satisfy the conditions above. Note that frequencies of the trigonometric functions are chosen carefully so that Hamiltonians are not periodic in time. Additionally, random matrices have the property of level repulsion, making it unlikely for the gap between the ground state and the first excited state to be anomalously small. 

Fig.~\ref{fig:Qs} compares the different $Q_D$'s suggested in Eq.~(\ref{Eq:Qds}) for a single Hamiltonian. It shows that the behavior is almost the same among the different proxies for computational complexity. Note that numerical verification confirms that other choices of random matrices exhibit the same behavior; therefore, additional figures are omitted for conciseness. For simplicity of presentation, we will use $Q_{D_1}$ as our $Q_D$ for the remainder of the paper.

\begin{figure}[t]
    \includegraphics[width=0.49\textwidth]{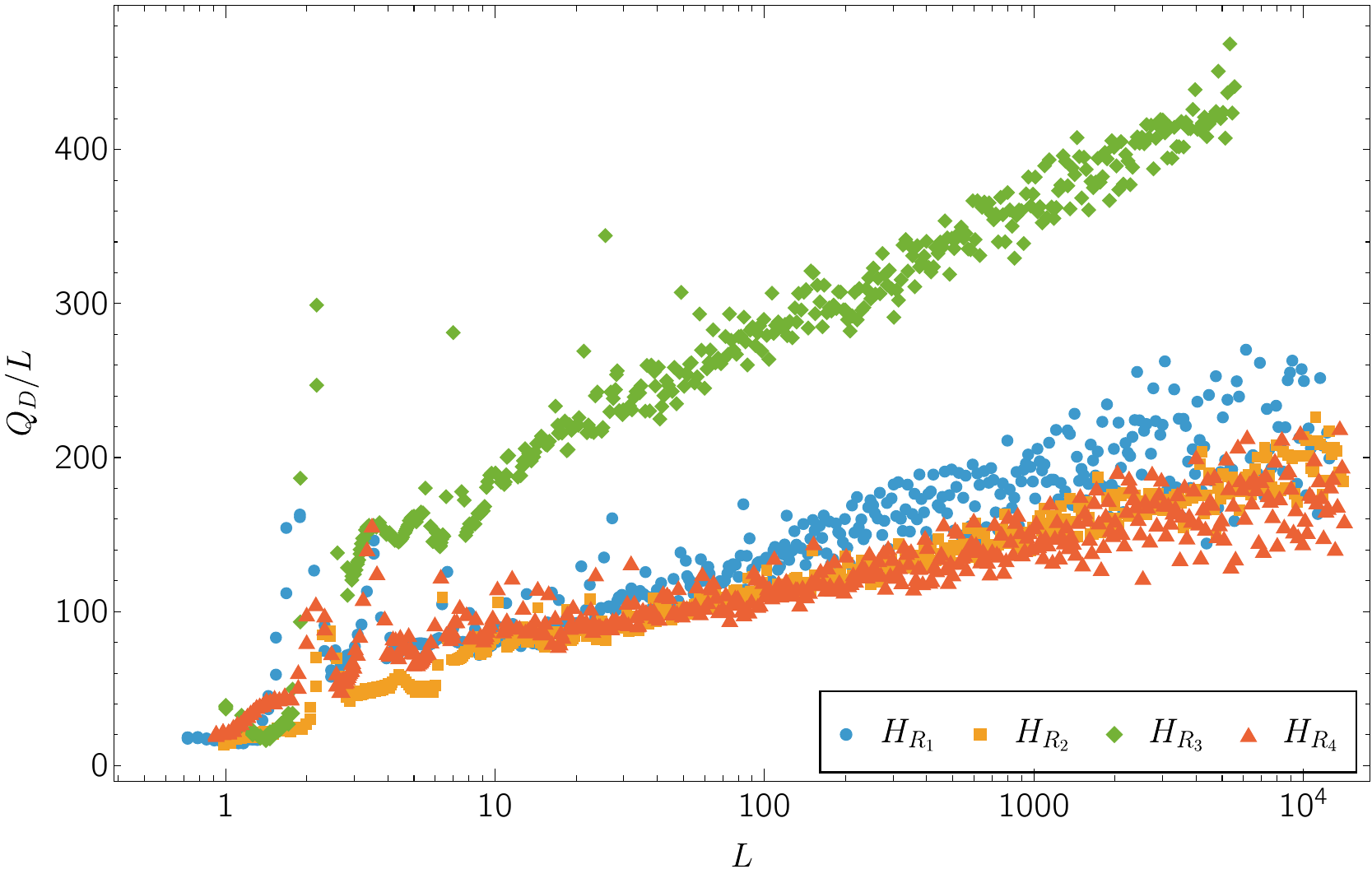}
    
    \caption{$Q_D$ is plotted with respect to the path length using four different choices of random Hamiltonians with the form of Eq.~(\ref{Eq:RandomH}). The threshold error $\epsilon_{\rm th}$ is chosen as 0.1. $R_i$ denotes the different random matrices for $H_1$ and $H_2$ in Eq.~(\ref{Eq:RandomH}).
    }

    \label{fig:QL}
\centering 
\end{figure}

In Fig.~\ref{fig:QL}, $Q_D$ is plotted with respect to $L$ using four different random Hamiltonians of this form. Note that the y-axis represents $Q_D/L$, and therefore $Q_D$ increases superlinearly with $L$ as one would expect if the conjecture in \cite{Cohen:2024nbk} is correct.  Note that there is nontrivial spread in the values of $Q_D/L$ showing that its value fluctuates with $L$, but that in all cases there is an overall secular increase with increasing $L$.  Due to these fluctuations, it is difficult to precisely isolate the behavior of secular growth, but it appears to be at least approximately linear from the plot, on which $L$ is given logarithmically. Thus, the asymptomatic growth of $Q_D$ is not inconsistent with $Q_D \propto L \log L$. However, regardless of the precise asymptotic form, the key finding is that growth of $Q_D$ is unambiguously superlinear in $L$, as conjectured.

\section{Discussion} \label{Sec:Discussion}

In the context of quantum computing, it is important to understand the scaling of the computational cost with respect to path length in the preparation of the ground state, as there are methods that are known to scale linearly~\cite{Cohen:2023dll}. Thus, for systems with long path lengths it is important to know whether adiabatic state preparation is competitive.

It is known that special cases exist in which the traversal of Hamiltonian paths with fixed errors has $Q_D$ growing linearly with $L$. Consider, for example, a Hamiltonian path of the form 
\begin{equation}
H(t) = \exp\left (-i \frac{v A \,t}{\sqrt{\langle g_0|A^2| g_0 \rangle}} \right ) H_0 \exp\left (i \frac{v A \,t}{\sqrt{\langle g_0|A^2| g_0 \rangle}} \right )  
\label{Eq:example}
\end{equation}
where $| g_0 \rangle$ is the ground state of $H_0$, $A$ is some Hermitian operator satisfying $\langle g_0|A| g_0 \rangle =0$ and $v$ is a constant velocity through which path is traversed. It is easy to show that with this construction the parameter $v$ corresponds to the velocity. An explicit example could be quantum mechanics in one dimension where $H_0=\frac{p^2}{2m} +V(x)$ and $A=p$. Such paths correspond to shifting the original Hamiltonian to an equivalent one with the same spectrum but at a different location. It is straightforward to show that for paths of this type, the error averaged over the path saturates to a value proportional to $v$  when the path is long and $v$ is small. Thus, the (average) error is independent of the path length at fixed $v$, while from Eq.~(\ref{Eq:QD1}) $Q_D$ grows linearly with $L$.  

However, such cases are really special and were constructed specifically to ensure linear scaling of the proxies for the computational cost. The key issue is what happens in generic cases. The examples in Section~\ref{Sec:Results} showed compelling evidence that the scaling was superlinear when the error was held fixed, albeit for very simple systems. Minimally this demonstrated that superlinear scaling occurs at least for some systems. Moreover, it is highly plausible that this superlinear behavior is the generic case and is almost certain to persist for Hamiltonians with much larger Hilbert spaces. In the context of quantum computing, there is no reason to believe that realistic cases of adiabatic evolution are likely to be special in the same way as the example in Eq.~(\ref{Eq:example}) and thus the proxies for the computational cost presumably scale typically superlinearly.

Superficially, this suggests that in the future, when confronted with very long paths, it would be sensible to choose a method with provably linear scaling in $L$ under fixed errors---such as the one proposed in Ref.~\cite{Cohen:2023dll}---over state preparation via adiabatic evolution. On the other hand, the simple examples studied here suggest that the scaling is only modestly superlinear, with the computational difficulty scaling consistently as $L \log L$. Thus, if the adiabatic approach has other computational advantages it is possible that it would be more efficient except in cases of exceptionally long paths. {\it A priori}, there is no reason to suppose that the adiabatic approach has such advantages. Indeed, it is possible that, apart from the modest advantage in the path length scaling, alternative methods may have other computational advantages. An obvious direction of future research is to discern under what circumstances, if any, is adiabatic quantum state preparation preferable when path lengths are long.

\begin{acknowledgments}

We would like to thank Maneesha Sushama Pradeep for helpful discussions. This work was supported in part by the U.S. Department of Energy, Office of Nuclear Physics under Award Number(s) DE-SC0021143, and DE-FG02-93ER40762.

\end{acknowledgments}

\bibliography{refs.bib}

\end{document}